\documentclass[reprint,amsmath,amssymb,braket,aps,prb,longbibliography]{revtex4-2} % PRX
\usepackage{graphicx,color,braket,amsmath,dcolumn,bm}% Include figure files
\usepackage[colorlinks=true,linkcolor=blue]{hyperref}
\newcommand{\corr}[1]{\textcolor{black}{#1}}

\begin{document}
\title{Testing the tomographic Fermi liquid hypothesis with high-order cyclotron resonance}

\author{Ilia Moiseenko}
\affiliation{Laboratory of 2d Materials for Optoelectonics, Moscow Institute of Physics and Technology, Dolgoprudny 141700, Russia}

\author{Erwin M\"{o}nch}
\affiliation{Terahertz Center, University of Regensburg, 93040 Regensburg, Germany}

\author{Kirill Kapralov}
\affiliation{Laboratory of 2d Materials for Optoelectonics, Moscow Institute of Physics and Technology, Dolgoprudny 141700, Russia}

\author{Denis Bandurin}
\affiliation{Department of Materials Science and Engineering, National University of Singapore, Singapore 117575}

\author{Sergey Ganichev}
\affiliation{Terahertz Center, University of Regensburg, 93040 Regensburg, Germany}

\author{Dmitry Svintsov}
\email{svintcov.da@mipt.ru}
\affiliation{Laboratory of 2d Materials for Optoelectonics, Moscow Institute of Physics and Technology, Dolgoprudny 141700, Russia}

\begin{abstract}

Recent theoretical studies of carrier-carrier scattering in degenerate two-dimensional systems have revealed radically different relaxation times for odd and even angular harmonics of distribution function. This theoretical concept, dubbed as 'tomographic Fermi  liquid', is yet challenging to test with dc electrical measurements as electron scattering weakly affects the electrical resistivity. Here, we show that linewidth and amplitude of electromagnetic absorption at the multiple harmonics of the cyclotron resonance carries all necessary information to test the tomographic Fermi liquid hypothesis. Namely, the  height and inverse width of $m$-th order cyclotron resonance ($m \ge 2$) is proportional to the lifetime of $m$-th angular harmonic of electron distribution function $\tau_m$, if probed at wavelengths exceeding the cyclotron radius $R_c$. Measurements of high-order cyclotron resonance at short wavelengths order of $R_c$ also enable a direct determination of all lifetimes $\tau_m$ from a simple linear system of equations that we hereby derive. Extraction of cyclotron resonance lifetimes from an experiment on terahertz photoconductivity in graphene shows that third-order resonance is systematically narrower than second-order one, supporting the prediction of tomographic Fermi liquid hypothesis.
\end{abstract}
\maketitle

An electronic excitation in a \corr{three-dimensional} Fermi liquid has a lifetime inversely proportional to the \corr{squared energy above the Fermi surface}, $\tau_{ee}\propto \delta \varepsilon^{-2}$~\cite{landau_Fermi_liquid}. This fact follows from strong phase-space restrictions for electron-electron (e-e) scattering events. It enables the introduction of weakly interacting quasiparticles in the system of strongly interacting electrons, forming the basis of Fermi liquid (FL) theory~\cite{Khalatnikov_Fermi_liquid}. Deviations from this scaling of electron lifetime are in the focus of modern condensed matter physics. Once the e-e scattering becomes strong, the conventional FL theory is no more able to predict the thermodynamic and kinetic properties. Situations favouring the breakdown of FL include reduced dimensionality~\cite{Tomonaga-Luttinger-state}, small Fermi surfaces~\cite{Abrikosov_possible_existence}, flat bands~\cite{Flat_bands_1,Flat_bands_2}, but are not limited to the latter.

The above unconventional examples implied stronger carrier scattering, as compared to the normal FL case. Recently, it was shown that electron scattering in 2d electron systems (2DES) may be weaker than in normal FL~\cite{Levitov_Tomographic}. More precisely, the odd angular harmonics of the distribution function $\delta f_m \propto e^{i m \theta}$ ($\theta$ is the momentum direction) were shown to relax via e-e scattering in an anomalously slow fashion. The scaling of odd-$m$ scattering rates follows a power-law dependence $\gamma_{m=2k+1} \equiv \tau^{-1}_{ee, m}\propto T^\alpha$ with $2<\alpha<4$~\cite{Kryhin2021,Hofmann_Lifetimes}. The even-$m$ harmonics relax in a normal fashion $\gamma_{m=2k} \propto T^2 \ln (\varepsilon_F/T)$~\cite{Zheng_Lifetime}, where $\varepsilon_F$ is the Fermi energy. \corr{The physics beyond such relaxation hierarchy comes from the large phase space for collisions with zero total momentum of incoming particles (head-on collisions) in two dimensions, as compared to three dimensions~\cite{Gurzhi_new_hydrodynamic}. Yet, head-on collisions do not alter the odd-$m$ harmonics due to their anti-symmetry with respect to momentum ${\bf p}$. Another type of collisions, the low-angle collisions, relax only the distributions with strong angular dependence, i.e. $m \gg 1$.}

This new regime of carrier transport in two dimensions, dubbed as 'tomographic Fermi liquid' (TFL), is challenging to test experimentally. The primary difficulty lies in the fact that e-e collisions do not directly affect electron mobility and resistivity in uniform fields and in large samples. They effect only on electron viscosity, the kinetic coefficient revealing itself only in highly non-uniform fields or small samples~\cite{Levitov_Tomographic,Kryhin_T_linear,Mollenkamp_Narrow_2deg}. As a result, the check of TFL hypothesis requires measurement of electron viscosity as a function of tuning parameters (e.g. sample width $W$~\cite{Levitov_Tomographic}, frequency $\omega$~\cite{Das_sarma_plasmons_TFL} or temperature $T$~\cite{Kryhin_T_linear}) and comparison of the obtained functional dependence with normal and tomographic FL theories. First, the measurement of viscosity is already challenging due to unavoidable residual scattering by impurities and phonons. Second, the NFL and TFL theories predict quite close scaling exponents for viscosity, which hinders the differentiation of these two regimes.

Here, we show that high-order cyclotron resonance (CR) in the 2D electron systems~\cite{MCR_old_exp,MCR_old_exp2,MCR_Old_exp_3,MCR_old_theor} carries unique and unambiguous information about relaxation times of angular harmonics of the distribution function. More precisely, the width of $m$-th cyclotron resonance $\Gamma_m$ and the relaxation rate of the $m$-th distribution function harmonic $\gamma_m$ are equal, $\gamma_m = \Gamma_m$. This relation holds for weak scattering, $\gamma_m \ll \omega_c$, $\omega_c = e|B|/m$ is the cyclotron frequency, and for relatively uniform ac fields, $qR_c \ll 1$, $R_c$ is the cyclotron radius. In highly non-uniform fields with $qR_c\sim 1$, a more general relation between $\Gamma_m$ and $\gamma_m$ emerges. The measurement of width for sequential cyclotron peaks enables testing whether even and odd distribution harmonics relax at different rates. 

Prior theoretical studies of CR linewidth %in the presence of disorder and electron interactions 
 based on diagrammatic techniques~\cite{Ando1976,Ting1977,Kallin1985} overlooked this relation. A possible reason is the necessary account for conservation laws upon e-e collisions, which amounts to quite a challenging evaluation of vertex functions. In the classical magnetic fields, solution of kinetic equation provides a more transparent approach to the problem. Prior kinetic theories of CR assumed  either identical relaxation rates for all angular harmonics~\cite{MCR_old_theor}, possibly excluding the relaxation of zeroth and first harmonics to account for conservation laws~\cite{Polini_Hall_viscosity,Kapralov2022a}, or were limited to the hydrodynamic approximation~\cite{Muller2008,cruise2024observability}. Suppression of the second CR harmonic by e-e collisions was evidenced in Ref.~\cite{CR_Ganichev_collisions}. Very recently, non-trivial modifications of cyclotron absorption due to the non-dissipative FL interactions were predicted~\cite{Afanasiev_Bernstein}. Here, we shall take a full account of different harmonic relaxation rates in the high-order CR problem.

The relation between $m$-th order CR linewidth and relaxation rate of $m$-th angular harmonic can be understood as follows. The electromagnetic field changes its direction $m$ times during the $m$-th order cyclotron resonance, causing an $m$-fold deformation of the Fermi surface, $\delta f(\theta_p) \propto e^{im\theta_p}$, see Fig.~\ref{fermisurface}. Of course, high-order cyclotron absorption can occur only in non-uniform ac fields. This non-uniformity adds a phase factor $e^{i q R_c \sin\theta_p}$ to the distribution function~(see \cite{pitaevskii2012physical,Kapralov2022a} and Appendix A). Still, this phase factor makes no effect on linewidth in the limit $qR_c\ll1$.

\begin{figure}[ht]
\center{\includegraphics[width=1\linewidth]{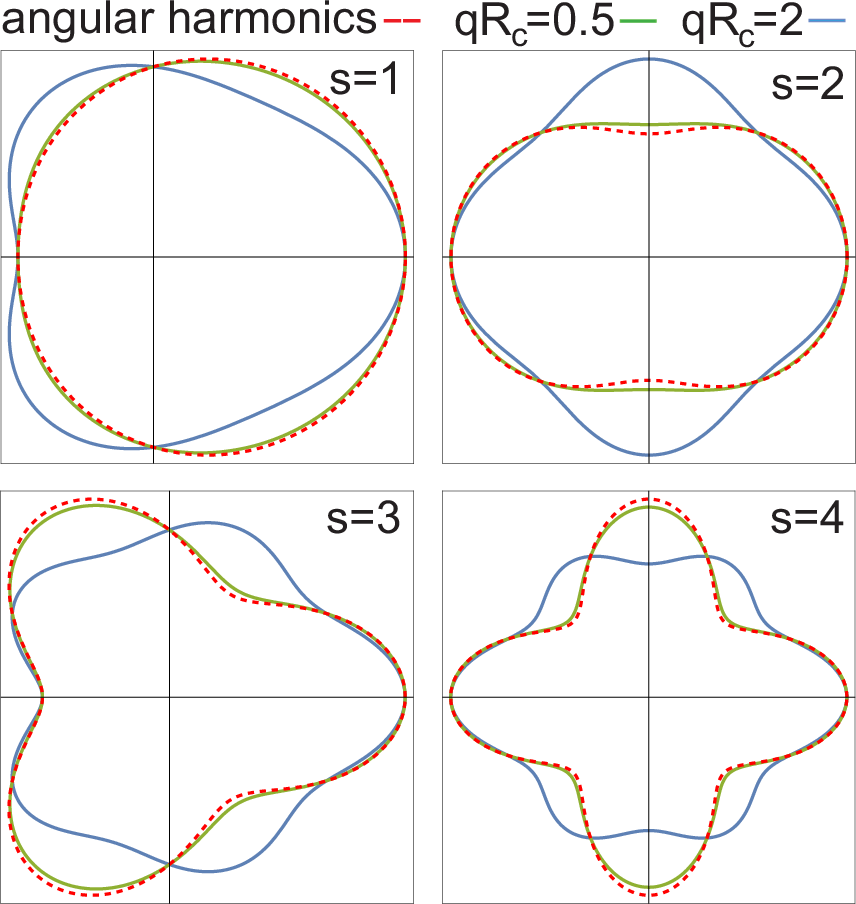}}
\caption{An arbitrary deformation of the Fermi surface can be presented as a superposition of angular harmonics. The deformed Fermi sphere under harmonic variation of Fermi momentum $p_f = p_0 + \delta p {\rm Re e^{i m \theta}}$ is shown with red dashed line for $m=1..4$. Green and blue lines show the deformations of Fermi sphere under the conditions of $m$-th order cyclotron resonance. Green line corresponds to almost uniform field ($qR_c = 0.5$), blue line -- to a non-uniform field with $qR_c = 2$.}
\label{fermisurface}
\end{figure}

We proceed to a rigorous solution of the CR problem in tomographic Fermi liquid. 
%To study the regime of tomographic hydrodynamics, we obtained an expression for the conductivity of the two-dimensional electron system (2DES) in a strong magnetic field. 
For this purpose, we solve the kinetic equation for an ac field-induced correction $\delta f$ to the distribution function:
\begin{multline}
\label{eq2}
-i\omega \delta f+iq v_F \cos \theta_p \delta f-e\frac{\partial {{f}_{0}}}{\partial {\bf p}}{\bf E} 
+\\
+{{\omega }_{c}}\frac{\partial  \delta f}{\partial \theta }=\mathcal{C}_{ee}\{ \delta f\}.
\end{multline} 
%\begin{equation}\label{eq1}
%  {\frac{\partial f}{\partial t}+\mathbf{V}\frac{\partial f}{\partial \mathbf{r}}-e(\mathbf{E}+\frac{1}{c}\mathbf{V}\times \mathbf{B})\frac{\partial f}{\partial \mathbf{p}}=\text{S}{{\text{t}}_{ee}}\{f\}},
%\end{equation}
Above, $\mathbf{q}$ is is the wave vector directed along the $x$-axis, ${\bf E} \propto \exp(-i\omega t+i\mathbf{qr})$ is the small ac electric field causing the CR directed, 
%$\mathbf{B}$ is the magnetic field perpendicular to the 2DES plane ($e_z$ $\Vert$ $B_z$), 
$v_F$ is the Fermi velocity in 2DES, $\theta$ is the angle between electron momentum and direction of field non-uniformity ${\bf q}$, and $\mathcal{C}_{ee}$ is the electron-electron collision integral.
% $f=f_{0}+\delta f\exp (-i\omega t+i\mathbf{qr})$, $f_0$ is the stationary distribution function of electrons in the 2DES, $\delta f$ is the small correction to the distribution function caused by the electric field, 
We further introduce the parametrization of distribution function in terms of angular harmonics
\begin{equation}
\label{eq-parametrization}
\delta f= \frac{\partial f_0}{\partial \varepsilon} \sum\limits_{m}{\chi_m e^{i m \theta_p }},\\
\end{equation}
The harmonic coefficients $\chi_m$ with the dimension of energy now weakly depend on $\varepsilon$, and the principal energy dependence is absorbed in the prefactor $\partial f_0/\partial \varepsilon$. The main convenience of representation (\ref{eq-parametrization}) is the simple structure of collision integral in the harmonics' basis, guaranteed by the rotational invariance:
\begin{equation}
{\mathcal C}_{ee}\{ \chi_m \}=-\gamma_{m} \chi_{m}.
\end{equation}
Above, we have introduced the relaxation rates for $m$-th angular harmonics of the distribution function $\gamma_m = 1/\tau_m$. The lowest three rates are zeros due to conservation of particle number and momentum upon collisions, $\gamma_{0,\pm 1}=0$~\cite{Khalatnikov_Fermi_liquid}. According to the tomographic Fermi liquid hypothesis, further harmonics satisfy $\{ \gamma_{2k} \equiv \gamma_{\rm even}\} \gg \{ \gamma_{2k+1} \equiv \gamma_{\rm odd}\}$.

\corr{Further on, it would be convenient to present the kinetic equation in the operator form similar to that used in quantum mechanics~\cite{Levitov_Tomographic,Levitov_AnnPhys}. We introduce the ket-vector for the distribution function $\ket{\chi}$ related to the angular harmonics $\ket{m}$ as $\ket{\chi} = \sum{\chi_m \ket{m}}$, and the ket-vector for the electric forces $\ket{F}$}
\corr{\begin{equation}
    \ket{F} = i \frac{ev_F}{\sqrt{2}}\left(E_+ \ket{1} +E_- \ket{-1} \right).
\end{equation}
Above, $E_\pm = (E_x \pm i E_y)/\sqrt{2}$ are the amplitudes of circularly polarized electric fields. With these notations, the kinetic equation becomes
\begin{equation}
\label{eq-cr-matrix}
   ( \omega {\hat I} - {\hat H} + i \hat{\mathcal{C}}_{ee}) \ket{ \chi} = \ket{F}.
\end{equation}
Here, $\hat I$ is the identity operator, the 'dynamic matrix' ${\hat H}$ governs the classical electron motion in the magnetic field and has the tridiagonal structure:
\begin{equation}
    \hat{H}=\left( \begin{matrix}
   ... & ... & 0 & 0 & 0  \\
   q{{v}_{F}}/2 & \left( m+1 \right){{\omega }_{c}} & q{{v}_{F}}/2 & 0 & 0  \\
   0 & q{{v}_{F}}/2 & m{{\omega }_{c}} & q{{v}_{F}}/2 & 0  \\
   0 & 0 & q{{v}_{F}}/2 & \left( m-1 \right){{\omega }_{c}} & q{{v}_{F}}/2  \\
   0 & 0 & 0 & ... & ...  \\
\end{matrix} \right),
\end{equation}
and $\hat{\mathcal{C}}_{ee} = {\rm diag}\{\gamma_m\}$ is the matrix representation of e-e collision integral.}

\corr{The solution of (\ref{eq-cr-matrix}) is reached once the eigen frequencies $\omega_s + i \Gamma_s$ and eigen vectors $\ket{s}$ of the dynamic operator ${\hat H} - i \hat{\mathcal{C}}_{ee}$ are found. The former correspond to the frequencies and linewidths of the $s$-th order cyclotron resonances. Performing the operator inversion, we find the distribution function:
\begin{equation}
    \ket{\chi} = \sum\limits_{s=-\infty}^{+\infty}{\frac{\braket{s|f}\ket{s}}{\omega - \omega_s - i \Gamma_s}},
\end{equation}
and the conductivity tensor
\begin{equation}
\label{eq-conductivity-symbolic}
    \sigma_{\alpha\beta} = \sigma_D \sum\limits_{s=-\infty}^{+\infty}{\frac{\omega}{\omega - \omega_s - i \Gamma_s}\braket{s|\alpha}\braket{\beta|s}}. 
\end{equation}
where $\sigma_D = i e^2 v_F^2 \rho(\varepsilon_F)/2\omega$ is the high-frequency Drude conductivity, $\rho(\varepsilon_F)$ is the density of states at the Fermi level. Equation (\ref{eq-conductivity-symbolic}) is applicable both to the 2d electron systems with parabolic bands (e.g. GaAs-based quantum wells) and to the massless electrons in graphene. It also applies both in the circular and Cartesian bases. In the first case, $\alpha$ and $\beta$ take on the values of $+1$ for the right-circular field and $-1$ for the left-circular field. In the second case, $\alpha = \{ x,y\}$ and $\beta = \{ x,y\}$ enumerate the Cartesian axes. The Cartesian eigenvectors are related to the angular harmonics via $\ket{x} = (\ket{+} + \ket{-})/\sqrt{2}$ and $\ket{y} = (\ket{+} - \ket{-})/\sqrt{2}i$. }

The calculated frequency-dependent conductivity of the tomographic Fermi liquid (\ref{eq-conductivity-symbolic}), is shown in Fig.~\ref{fig-sigma-mcr}~\footnote{Application of Eq.~\ref{eq-conductivity-symbolic} requires the knowledge of eigenvectors $\ket{s}$ and eigenvalues $\omega_s + i \Gamma_s$ for the dynamic matrix. For arbitrary scattering strength, they can be found numerically by retaining a large (but finite) number of angular harmonics $m_{\max}$. The computational complexity here scales linearly with $m_{\max}$ due to the tridiagonal structure of the matrix. In the weak-scattering limit, the perturbative results (\ref{eq-crs-finite-q},\ref{eq-gammas-finite-q}) apply.}. The real part of the conductivity displays sharp resonances at $\omega = m \omega_c$, $m \ge 1$. The cyclotron harmonics become stronger with increasing the field non-uniformity, i.e. at larger $q$. The main CR broadens at larger $q$, which is a consequence of the viscous character of damping in FL. So far, all these observations are in agreement with previous studies of CR in 2DES with weak carrier-carrier collisions~\cite{Kapralov2022a,Polini_Hall_viscosity,Alekseev_viscoelastic_resonance}.

\begin{figure}[ht]
\center{\includegraphics[width=1\linewidth]{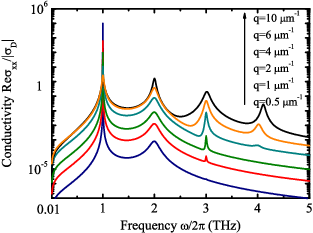}}
\caption{The real part of the longitudinal 2DES conductivity Re$\sigma_{xx}$, normalized by the collisionless Drude conductivity $\sigma_{\text{D}}=\frac{n{{e}^{2}}}{m\omega}$ for different values of $q$ with $\tau_{even}$=2 ps, $\tau_{odd}$=10 ps, $\omega_{c}/2\pi$=1 THz}
\label{fig-sigma-mcr}
\end{figure}

A distinctive feature of tomographic electron fluid is the relation between widths of subsequent cyclotron resonances. Namely, the third CR in Fig.~\ref{fig-sigma-mcr} is very narrow compared to the second one, while the fourth is broad again. This alternating character of the resonance widths is a direct consequence of slow decay of the odd distribution harmonics and fast decay of even ones. The above rule holds as soon as the field remains uniform, $qR_c \ll 1$. For a cyclotron frequency $\omega_c/2\pi = 1$ THz and Fermi velocity $v_F = 10^{6}$ m/s, the cyclotron radius is estimated as $R_c \approx 0.15$ $\mu$m. In highly non-uniform fields (black curve at Fig.~\ref{fig-sigma-mcr}), seemingly all harmonics are equally broad, though the situation is even more intricate.

\begin{figure}[ht]
\center{\includegraphics[width=1\linewidth]{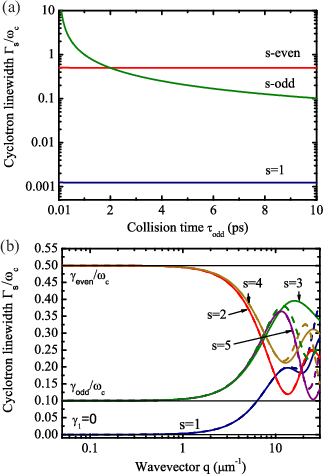}}
\caption{Cyclotron linewidths $\Gamma_s$ in the tomographic Fermi liquid. (a) $\Gamma_s$ in units of $\omega_c$ vs variable relaxation time of odd angular harmonics $\tau_{odd}$ with constant $\tau_{even}$=2 ps. Wave vector  $q=1$ $\mu$m$^{-1}$. (b) Linewidth $\Gamma_s$ for several lowest $s=1...5$ vs wave vector of the electromagnetic field $q$. The results obtained by numerical diagonalization of dynamic matrix are shown with solid lines, the analytical result of the perturbation theory is shown by the dashed ones. In both panels, charge carrier concentration is $n=10^{12}$ cm$^{-2}$, cyclotron frequency $\omega_c/2\pi = 1$ THz}
\label{fig_linewidth}
\end{figure}

As apparent from conductivity spectra, the width of CR is the quantitiy affected by the tomographic nature of the 2d Fermi liquid. A direct measure of to this width is the imaginary part of eigen frequency $\Gamma_s$. We now proceed to its detailed studies. In Fig.~\ref{fig_linewidth} (a), we prove numerically that variations of the odd angular harmonic lifetimes $\gamma_{2k+1}$ affect only the width of odd-$s$ cyclotron resonances in weakly non-uniform fields $qR_c \ll 1$. We use identical scattering rates for all even and all odd harmonics and wave vector $q=1$ $\mu$m. We indeed observe that all odd and even $\Gamma_s$ coalesce onto two curves. The linewidths $\Gamma_{2k}$ are not affected by variations of $\tau_{2k+1}$ at  all. Small but finite linewidth of the $s=1$ resonance is due to the finite value of wave vector and thus the presence of viscous damping.

The numerically observed patterns in $\Gamma_s$ gain a simple explanation in the framework of perturbation theory for the operator ${\hat H} - i \hat{\mathcal{C}}_{ee}$, considering the collisions as small perturbation. From mathematical viewpoint, ${\hat H} $ is equivalent to the Hamiltonian of the tight-binding chain in dc electric field, $\omega_c$ playing the role of voltage drop along one cell, and $qv_F/2$ playing the role of hopping integral. The eigenvalues of ${\hat H}$ are perfectly localized each at $s$-th harmonic (atomic site) in the absence of spatial dispersion (hopping). This implies the identity between cyclotron resonances $\ket{s}$ and angular harmonics $\ket{m}$ in the limit $qR_c \ll 1$. In this limit, inclusion of collisions trivially adds the damping $\Gamma_s = \gamma_s$ to each eigenfrequency.

In the presence of spatial dispersion, the eigenvectors of ${\hat H}$ are spread across various angular harmonics (atomic sites) according to~\cite{Saitoh1973}:
\begin{equation}
\label{eq-crs-finite-q}
   \braket{m|s} = J_{|s-m|}(qR_c),
\end{equation}
where $J_l(x)$ is the Bessel function of the $l$-th order. The first-order perturbative correction to the frequency of the spatially-dispersive states (\ref{eq-crs-finite-q}) is purely imaginary,
\begin{gather}
\label{eq-gammas-finite-q}
    \delta \omega = \bra{s} i \hat{\mathcal{C}}_{ee} \ket{s} = i \Gamma_s,\\
    \Gamma_s = J_{0}(q{{R}_{c}})^{2}{{\gamma }_{s}}+\sum\limits_{i=1}^{\infty }{{{J}_{i}}{{(q{{R}_{c}})}^{2}}(}{{\gamma }_{s+i}}+{{\gamma }_{s-i}})
\end{gather}
The result of perturbation theory (\ref{eq-gammas-finite-q}) reproduces very well the numerically obtained linewidths $\Gamma_s$ up to quite a large $qR_c \sim 10$, as shown in Fig.~\ref{fig_linewidth} (b). One observes that odd linewidths, being initially small, start growing at $qR_c \sim 1$. This occurs due to the 'admixture' of the even angular harmonics to the odd cyclotron resonance. Remarkably, the linewidths of the even CRs (say, 2nd and 4th) can drop down to the very small values. This occurs due to the oscillatory nature of the spectral weight $|J_0(qR_c)|^2$, and happens at $qR_c$ equal to the zeros of Bessel function $J_0$. We note finally that an expression for the viscous damping of the principal CR~\cite{Alekseev_viscoelastic_resonance,Polini_Hall_viscosity,Kapralov2022a} immediately follows from a more general equation (\ref{eq-gammas-finite-q}).

\begin{figure}[ht]
\center{\includegraphics[width=1\linewidth]{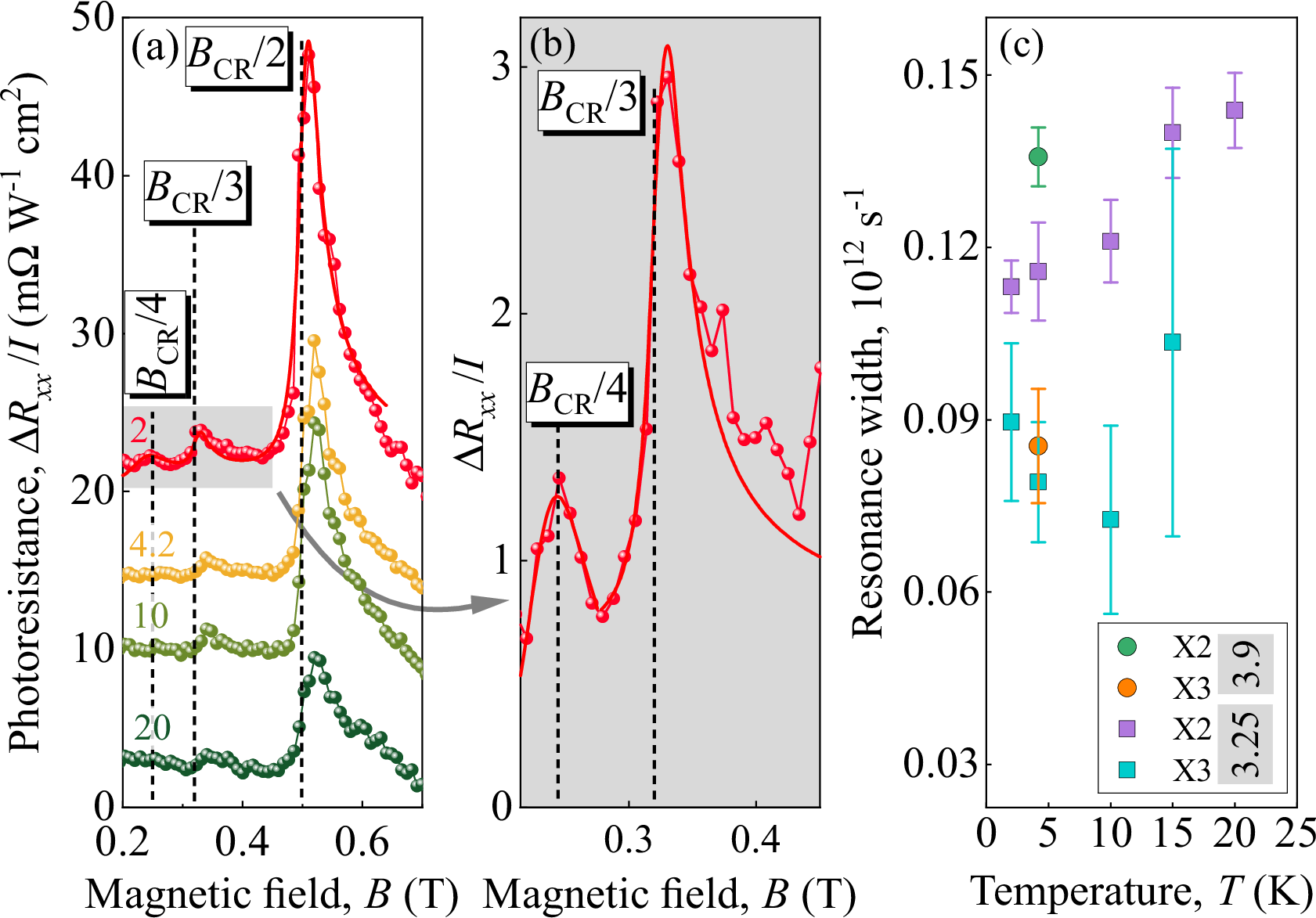}}
\caption{
%(a)  Graphene photoresistance (PR)  as a function of magnetic field for different values of temperature. Radiation frequency and carrier density are $\omega/2\pi$=0.69~THz and $n_{e} = 3.9 \times$10$^{12}$~cm$^{-2}$, respectively. The PR is defined as a change in the longitudinal resistance upon illumination $\Delta R_{xx}$, normalized by radiation intensity $I$. Dashed vertical lines mark calculated positions of the CR harmonics. (b) A blow-up of the gray shaded area in panel (a) highlighting the 3rd and the 4th CR's harmonics.  Black curves in both panels show fits of CR harmonics shape after Eq. (\ref{eq-approximation}). The fits are exemplary illustrated for the lowest temperature $T = 2$~K. (c) Extracted widths of the 2nd (purple points) and 3rd (cyan points) CR harmonics widths as a function of temperature.
\corr{Measured graphene photoresistance (PR) as a function of magnetic field for different values of temperature. The PR is defined as a change in a longitudinal resistance upon illumination $\Delta R_{xx}$, normalized by radiation intensity $I$.
Panel (a) shows the $B$-dependent PR in the vicinity of 2nd, 3rd and 4th CR harmmonics at different temperatures (marked by numbers), panel (b) shows the magnified view of PR at $T=2$ K in the vicinity of 3rd and 4th harmonics. Solid lines show the fit according to Eq.~(\ref{eq-approximation}), dashed vertical lines label the theoretically anticipated position of the CR. Radiation frequency is $f=0.69$ THz; carrier density is $3.9\times$10$^{12}$~cm$^{-2}$ (c) Extracted widths of the 2nd (X2) and 3rd (X3) cyclotron resonances at different temperature. Green and orange data points were obtained at a different carrier density $n_s=3.25\times10^{12}$ cm$^{-2}$ and $T=4$ K (raw data in Appendix D)}
}
\label{fig_experiment}
\end{figure}

The experimental test of the TFL hypothesis now amounts to the measurement of the 2d conductivity at a given wave vector and frequency $\sigma(q,\omega)$. Its real part governs the absorbed power $P_{\rm abs} = 1/2 \sum_{q}{\sigma'(q,\omega) |E_{q\omega}|^2}$, where $E_{q\omega}$ is the spectral composition of the electric ac field in the 2DES plane. One method for measuring $P_{\rm abs}$ relies on transmission spectroscopy of grating-gated 2DES~\cite{MCR_old_theor}. Such data are available only for the 2DES with modulated lateral doping~\cite{MCR_old_exp,MCR_old_exp2}, which complicates their analysis.

Another powerful technique for measuring the absorption spectra of low-dimensional systems relies on the photoresistance~\cite{Muravev_Photoconductivity,Ganichev_Photoconductivity}, i.e. the change in the $dc$ resistance $\Delta R_{\rm xx}$ induced by radiation.  It relies on the fact that $\Delta R_{\rm xx}$ is proportional to the absorbed electromagnetic power, and resonances in absorption at certain $\omega$ and $B$ should greatly enhance the measured value of $\Delta R_{\rm xx}$.

\corr{We have performed the photoresistance measurements on a high-quality graphene sample in classically strong magnetic fields $B<1$ T. The CR and its harmonics were excited by coherent THz radiation generated by a continuous wave optically pumped molecular laser. Details on sample fabrication and experimental technique can be found in Appendices B and C. We intentionally fabricated narrow ($\sim1~\mu$m in  width) metallic Hall contacts to the sample partially embedding into the channel. This configuration provides highly non-uniform electromagnetic fields and is thus particularly suitable for observation of high-order CRs~\cite{Alonso-Gonzalez2014}.}

\corr{Figure~\ref{fig_experiment} (a,b) shows the example of the measured photoresistance at $f=0.69$ THz and  $n_s= 3.9\times$10$^{12}$~cm$^{-2}$, where up to three CR overtones ($s=2...4$) are observed. We note that the PR signal at the main $s=1$ CR is very weak at that frequency and density, but re-appears at larger $f$ and $n_s$ (Appendix D)~\footnote{A possible reason for CR weakening at large $n_s$ lies in radiative damping\cite{Muravev_CR_fine}, i.e. screening of the incident radiation by highly conductive 2DES.}. At given $n_s$ controlled by the back gate, we confidently identify all spikes to CR or its harmonics according to $B = 2\pi m f/|e|s$, where $s$ is an integer and $m = \hbar\sqrt{\pi n_s}/v_F$ for single-layer graphene. The anticipated resonance positions are marked by vertical dashed lines in Fig.~\ref{fig_experiment} (a-b), and agree well with observed photoresistance peaks. This agreement enables further linewidth analysis.
}

To reproduce the highly asymmetric shape of the absorption, one has to account for the plasmonic effects, i.e. screening of the incident field $E_0$ by the 2d electrons~\cite{Muravev_CR_fine}. They are taken into account by relating the total field $E_{q\omega}$ to the incident field $E_0$ via the dielectric function of 2DES $\epsilon(q,\omega)$, $E_{q\omega} = E_0/\epsilon(q,\omega)$. The dominant contribution to the anti-screening comes from the waves with nearly-zero group velocity, the so-called Bernstein modes~\cite{Dai_CR_spike,Dorozhkin2021,Zabolotnykh_BMs}. In such situation, the spectral dependence of absorbed power is suitably described by~\cite{Bandurin2022a} 
\begin{equation}
\label{eq-approximation}
P_{\rm abs}={{P}_{0}} + \sum\limits_{s=2}^{\infty}{A_s \operatorname{Re}\left( \frac{1}{\sqrt{\omega_s^*-\omega +\text{i}\Gamma_s }} \right)}.
\end{equation}
%Expression (\ref{eq-approximation}) describes a highly asymmetric spectrum with a long shoulder at frequencies below $s\omega_c$, and an abrupt drop at frequencies above $s \omega_c$. It matches well the data on terahertz-induced photoconductivity in graphene in classically strong magnetic fields $B \lesssim 1$ T~\cite{Bandurin2022a}.

We extracted the linewidths $\Gamma_s$ of cyclotron resonances by fitted the photoresistnce data with model (\ref{eq-approximation}) considering $A_s$, $\omega^*_s$ and $\Gamma_s$ as fitting parameters. The extracted values of $\Gamma_s$ together with their error bars are shown in Fig.~\ref{fig_experiment} (c). The fitting procedure shows that third-order CR is systematically narrower than second-order one at all temperatures where it is seen, $2$ K $<T < 20$ K. This fact agrees with TFL hypothesis qualitatively. Namely, relaxation of the odd-order (third) cyclotron resonance appears to be weaker than that of the even order (second). The theory outlined above traces this fact to the different relaxation rates of even- and odd-order distribution functions.

Small scattering rate of the third harmonic of distribution function, as compared to the second one, cannot be explained within impurity or electron-phonon scattering models. It is easy to show that both these models predict stronger relaxation with increasing the harmonic number $m$ (see \cite{Baker_Viscous_Ballistic} and Appendix E). The TFL hypothesis is currently the only one capable of explaining the unusual relation $\gamma_3 < \gamma_2$. Still, it is worth noting that the $T$-dependence of relaxation time extracted from the experiment is not a power-law one. At least, the relaxation rates $\gamma_{2,3}$ have a large $T$-independent background which can be attributed to the built-in disorder.

To conclude, we have shown that the width of $m$-th order cyclotron resonance in a two-dimensional system is linked to the relaxation rate of the $m$-th angular harmonic of the distribution function. For long-wavelength fields these two quantities are exactly equal. The magnetoabsorption data of high-quality graphene at THz frequencies indicate on weaker relaxation of the third order CR, as compared to the second order one. Such anomalous relation between relaxation rates of the 2nd and 3rd angular harmonics of distribution function points to the validity of tomographic Fermi liquid hypothesis. Higher-resolution measurements are further required to test the anomalous scaling of relaxation rates with temperature.

%We would like to conclude our discussion by describing the optimal experimental conditions at which multiple cyclotron resonances carry the information about the details of electron-electron scattering. First, the odd scattering rates should be higher than momentum-relaxing impurity scattering. By order-of-magnitude, this results in a limitation to temperature $T$ and carrier mobility $\mu$ of the form $\hbar^{-1}\varepsilon_F (T/\varepsilon_F)^\alpha > e / (m^* \mu)$. Second, the ac electric field should be largely non-uniform, such that high-order Fourier harmonics of electric field could couple to the multiple cyclotron resonances. Formally, this results in $q v_F/\omega_c \gtrsim 1$, where $q$ is limited from above by the inverse size of the smallest scatterer in the system $q \lesssim 1/d$. Taking $\omega_c/2\pi = 1$ THz and $v_F = 10^6$ m/s, we arrive at $d < 150$ nm. This condition can be achieved in 2DES with keen metallic contacts or covered by metallic nanostructures with sharp edges where, strictly speaking, $q$ has no upper limit.

The work of I.M., K.K. and D.S. was supported by the Ministry of Science and Higher Education of the Russian Federation (grant No. FSMG-2021-0005). S.D.G. and E.M. acknowledge the financial support of the Deutsche Forschungsgemeinschaft (DFG, German Research Foundation) via Project-ID 314695032 – SFB 1277 (Subproject A04) and of the Volkswagen Stiftung Program (97738). D.A.B. acknowledges the support from MOE AcRF Tier 1 grant (\#22-5390-P0001).

\appendix
\begin{widetext}

\section{Solution for cyclotron resonance in collisionless 2d electron system}
We recall here some basic information about the kinetics of 2d electrons under cyclotron resonance and in the absence of collisions. The distribution function $\delta f(\theta)$ in this problem is presented as~\cite{pitaevskii2012physical,Kapralov2022a}
\begin{equation}
	\label{Eq-cyclotrn-expansion}
	\delta f (\theta) = e^{i q R_c \sin\theta}\sum\limits_{s=-\infty}^{+\infty}{\frac{g_s\{ {\bf E}\}}{\omega+i\delta -s\omega_c} e^{i s \theta}},    
\end{equation}
where $R_c = v_F/\omega_c$ is the cyclotron radius, and the functions $g_s\{ {\bf E}\}$ depend linearly on ac electric field and smoothly (non-resonantly) -- on dc magnetic field and frequency:
\begin{equation}
	\label{Eq-collisionless-df}
	g_s = \frac{\partial f_0}{\partial p} \left[ is \frac{J_s(qR_c)}{qR_c} e E_x + J_s'(qR_c) eE_y\right]\frac{1}{\omega+i\delta -s\omega_c}.
\end{equation}
Each $s$-th term in the sum (\ref{Eq-cyclotrn-expansion}) is resonantly excited when $\omega = s \omega_c$, which enables its interpretation as $s$-th order cyclotron resonance. The $\theta$-dependence of the distribution function at the exact $s$-th cyclotron resonance is almost harmonic one, $\delta f \approx g_s e^{i s\theta}$. The coincidence is exact in the long-wavelength limit $q R_c \ll 1$. Under this condition, the exponential prefactor $e^{iq R_c \sin \theta}$ associated with field non-uniformity can be neglected in Eq.~\ref{Eq-cyclotrn-expansion}. This fact is illustrated in Fig.~1 of the main text.  Plotting the Fermi surface deformations under the conditions of $s$-th order cyclotron resonance (green and blue) along with purely harmonic variations (red), we see that they are almost indistinguishable up to $qR_c \sim 1$.

Equations (9) and (11) of the main text could be alternatively obtained by recalling the Fourier expansion of the 'phase modulated signal' $e^{i q R_c \sin\theta}$ which enters the distribution function (\ref{Eq-cyclotrn-expansion}):
\begin{equation}
	\label{eq-phase-modulation}
	e^{i q R_c \sin\theta_p} = \sum\limits_{m=-\infty}^{+\infty}{J_m(qR_c)e^{im\theta_p}}.
\end{equation}
According to Eq.~(\ref{eq-phase-modulation}), the 'side harmonics' of the distribution function under high-order cyclotron resonance decay as $J_m(qR_c)$.

\section{Sample fabrication}

The encapsulated monolayer graphene samples used in this work were prepared by a hot-release method described in Ref.~\cite{Purdie2018}. First, the graphene flakes were mechanically exfoliated from a high-purity pyrolytic graphite crystal using the scotch-tape technique~\cite{Novoselov2004}. Layers of hexagonal boron nitride (hBN) were used to encapsulate the graphene flake to protect it from the surrounding. The resulting hBN/MLG/hBN van der Waals structure was stacked on a conventional p$^{++}$-doped Si/SiO$_2$ silicon wafer with 285~nm SiO$_2$ thickness. The structure was then processed with electron beam lithography to form the contact regions. An additional etch mask was used to process the special contact geometry. Figure~\ref{Fig1}(a) shows a schematic cross-section and Fig.~\ref{Fig1}(b) shows photomicrographs. More information can be found in Ref.~\cite{Bandurin2022a}.
\begin{figure}[ht]
	\centering \includegraphics[width=1.0\linewidth]{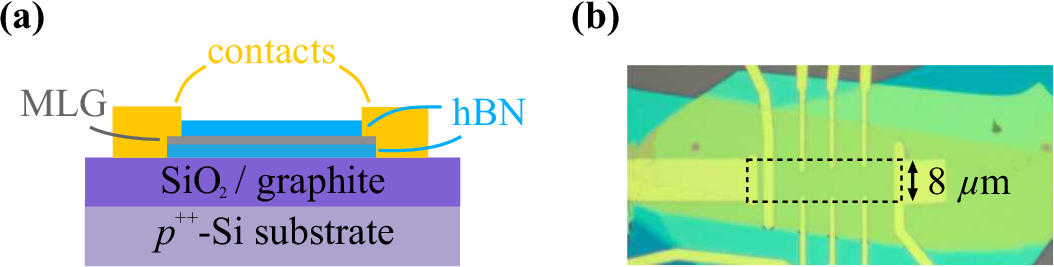}
	\caption{Cross section of the graphene structure. (b) Photomicrograph of the device with special contacts protruding into the Hall channel.
	}
	\label{Fig1} 
\end{figure}

\section{Experimental Technique}

In our experiments the coherent terahertz (THz) radiation is generated by a 	continuous wave terahertz molecular gas laser optically pumped by a CO$_2$ laser. We used methanol, difluoromethane and 	formic acid as active media to produce linearly polarized laser radiation 	with frequencies of $f = 2.54, 1.63$ and 0.69~THz [photon energies $\hbar\omega = 10.5, 6.74$ and 2.85~meV]. The radiation power lies in the range of 20 to 80~mW at the sample's position depending on the radiation frequency. The FWHM spot diameters at the sample were $d = 1.5$ [$f=2.54$~THz], 1.75 [$f=1.63$~THz] and 3.0~mm [$f=0.69$~THz]. The spot shape was controlled by a pyroelectric camera.

The photoconductivity/photoresistivity has been measured applying two techniques:  (i) applied a direct current with positive (negative) direction  to the sample and (ii) the double 	modulation setup~\cite{Kozlov2011,Otteneder2018,Hubmann2023}. In the former, a direct current is applied to the sample. The chopped THz radiation induces a voltage drop across a load resistor, which is then measured by the standard lock-in technique using the chopper frequency as the reference signal. Repeating it for both polarities of the applied direct current yields the photoconductivity (photovoltage) response by subtracting (adding) the signals obtained for positive and negative polarities.

The double modulation setup is an alternative and more sensitive method, which was used to obtain most of the data sets shown in this work. Here, instead of a direct current, an alternating current is applied to the sample. The frequency of the current ($f_{\rm ac} = 7.757$~Hz) should be much smaller than the chopper frequency ($f_{\rm chop} = 140$~Hz) modulating the THz radiation. To obtain the photoconductivity, two lock-in amplifiers are connected in series, the first of which is locked to $f_{\rm chop}$ and the second to $f_{\rm ac}$: the first lock-in filters out the dark signal, which then consists of a constant component that is proportional to the photovoltage, and a component that is modulated with the frequency $f_{\rm ac}$ being proportional to the photoconductivity. This signal is fed into the second lock-in, which is locked to $f_{\rm ac}$ and provides a voltage determined by the magnitude of the photoconductivity.

\section{Additional experimental data}

\begin{figure*}[h]
	\centering\includegraphics[width=1.0\linewidth]{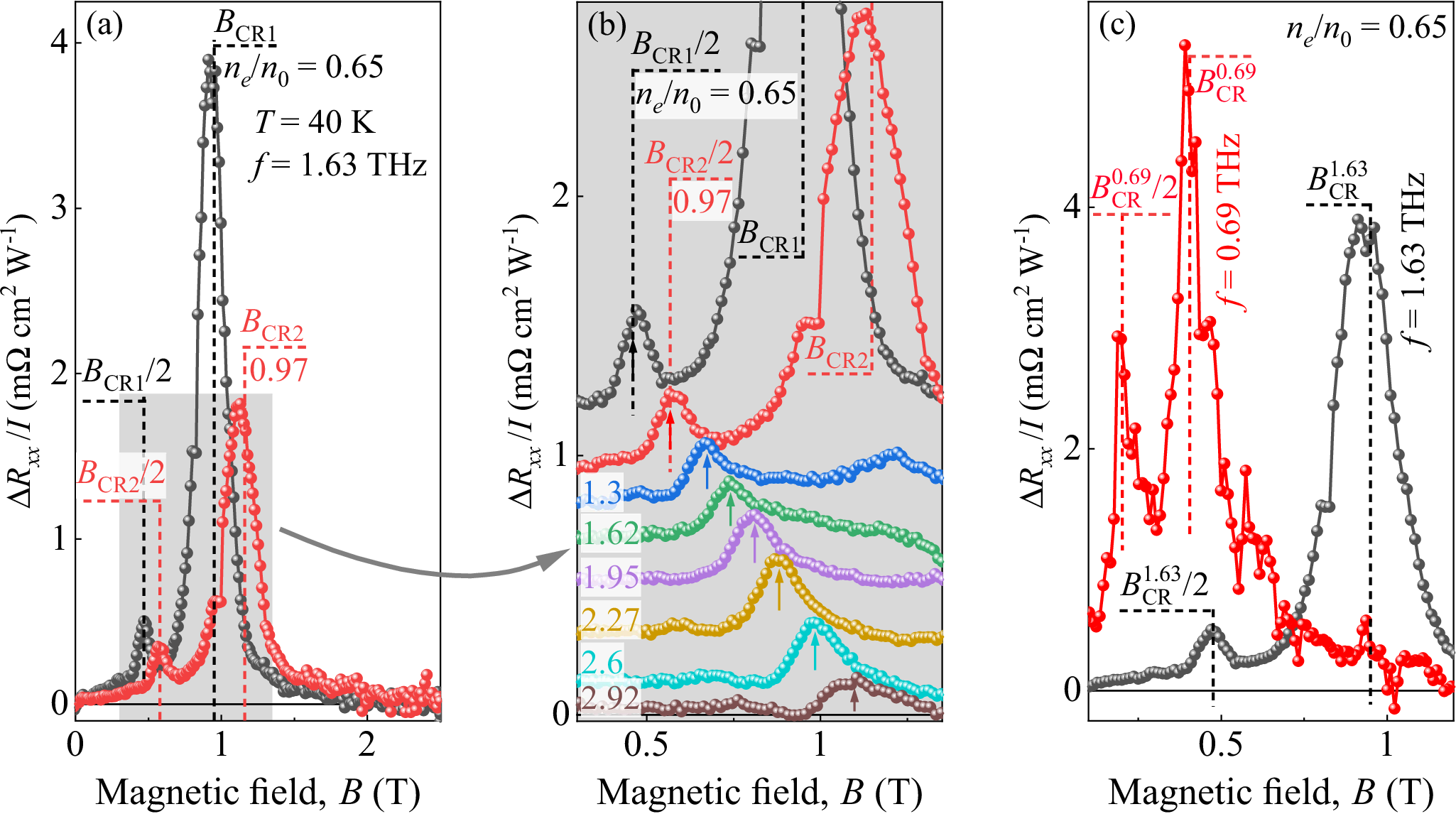}
	\caption{Graphene photoresistance as a function of magnetic field for different values of carrier concentration labeled by numbers in units of $n_0 = 10^{12}$~cm$^{-2}$. The photoresistance is defined as a change in the longitudinal resistance upon illumination $\Delta R_{xx}$, normalized by radiation intensity $I$. Dashed vertical lines mark the position of the CR ($B_{\rm CRi}$) and its second harmonic ($B_{\rm CRi}/2)$ labeled correspondingly. Here $i$ is integer.  Panel  (a) shows the data for the radiation with $f=1.63$~THz and two carrier densities.Vertical dashed lines shows the position of the calculated CR magnetic fields and its harmonics.  Panel (b) shows a blow-up of a set of curves for the whole range of carrier densities investigated. Panel (c) shows a comparison of the photoresistance for $f = 0.69$ (red) and 1.63~THz (black) for a similar carrier density of $n_e = 0.65\times10^{12}$~cm$^{-2}$. The radiation frequency and temperature used here are $\omega/2\pi = 1.63$~THz and $T = 40$~K, respectively.}
	\label{fig_experiment_184}
\end{figure*}

Figures~\ref{fig_experiment_184} and \ref{fig_experiment_118} show data sets of the photoresistance in graphene normalized to the corresponding radiation intensity measured with higher radiation frequencies $f = 1.63$ and 2.54~THz, respectively. The temperature was kept at $T = 40$~K to suppress the emergence of the Shubnikov-de Haas-like oscillations in the photoresistance. The data sets in both figures show pronounced resonant responses for higher and lower magnetic fields. The former, which is illustrated in full scale in Figs.~\ref{fig_experiment_184} (a, c) and \ref{fig_experiment_118} (a), can be clearly attributed to the main cyclotron resonance (CR), whose position fits to the calculated one (given by correspondingly colored vertical dashed lines), using the sample's carrier density and a Fermi velocity of $v_{\rm F} = 10^6$~m/s. The CR position scales with $B_{\rm CR} = 2\pi\hbar f\sqrt{\pi n_e}/ev_{\rm F}$ and the CR shape becomes broader with higher carrier concentration due to the effect of radiation screening~\cite{Muravev_CR_fine}. Figure~\ref{fig_experiment_184}(c) also shows the CR  obtained at low carrier density $n_e =0.65\times10^{12}$~cm$^{-2}$ applying radiation frequencies of $f= 1.63$ and 0.69~THz.

\begin{figure*}[ht]
	\centering\includegraphics[width=0.71\linewidth]{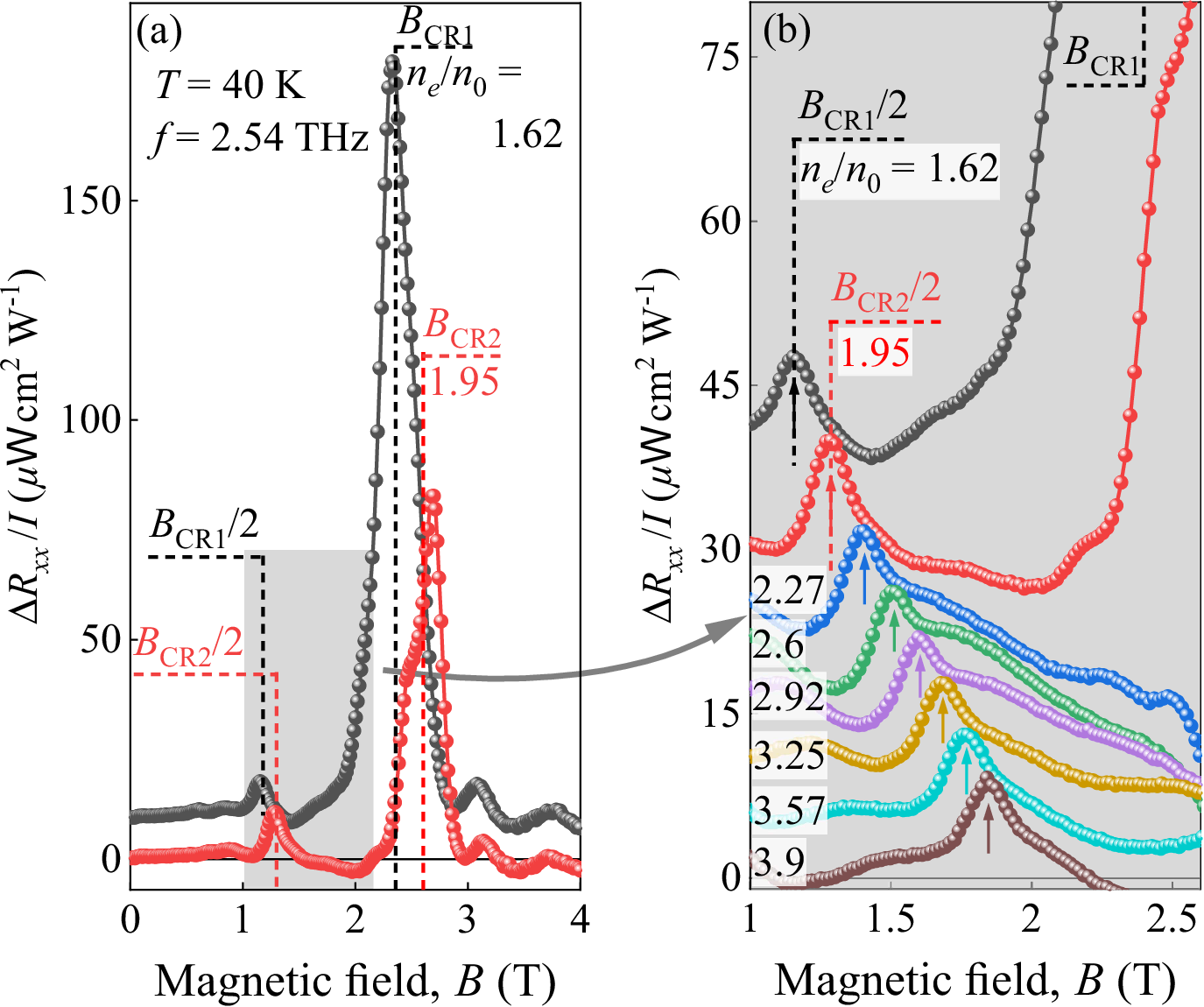}
	\caption{Graphene photoresistance vs magnetic field at illumination frequency $f=2.54$ THz. (a) Photoresistance curves at two characteristic values of carrier concentration $n_s = 1.62\times10^{12}$ cm$^{-2}$ (black) and $n_s = 1.95\times10^{12}$ cm$^{-2}$ (red). Dashed vertical lines mark up the positions of cyclotron resonances expected from quasi-classical theory (b) Zoom-in of the photoresistance curve in the vicinity of $s=2$ cyclotron resonance, with extra curves at other carrier densities from $n_s = 1.62\times10^{12}$ cm$^{-2}$ (black) to $n_s = 3.9\times10^{12}$ cm$^{-2}$ (brown)
	}
	\label{fig_experiment_118}
\end{figure*}

At lower magnetic fields we also observe a resonance-like behavior. Although somewhat weaker than the main CR, this peaks emerge in the vicinity of the second CR harmonic, $B_{\rm CR}/2$. Figures~\ref{fig_experiment_184} (b) and \ref{fig_experiment_118} (b) display the evolution of the position of the second CR harmonic with the carrier density, where the colored arrows label their calculated positions. Note that at high carrier densities the CR signal in response to the radiation with $f=0.69$~THz is rather weak whereas its harmonics have large magnitude, much larger than the CR one. This is the case of the data for $n_{e} = 3.9 \times$10$^{12}$~cm$^{-2}$ presented in the main text. Where the CR signal at $B_{\rm CR}$ is at least ten times smaller than the one for the second harmonic.

\begin{figure*}[ht]
	\centering\includegraphics[width=0.9\textwidth]{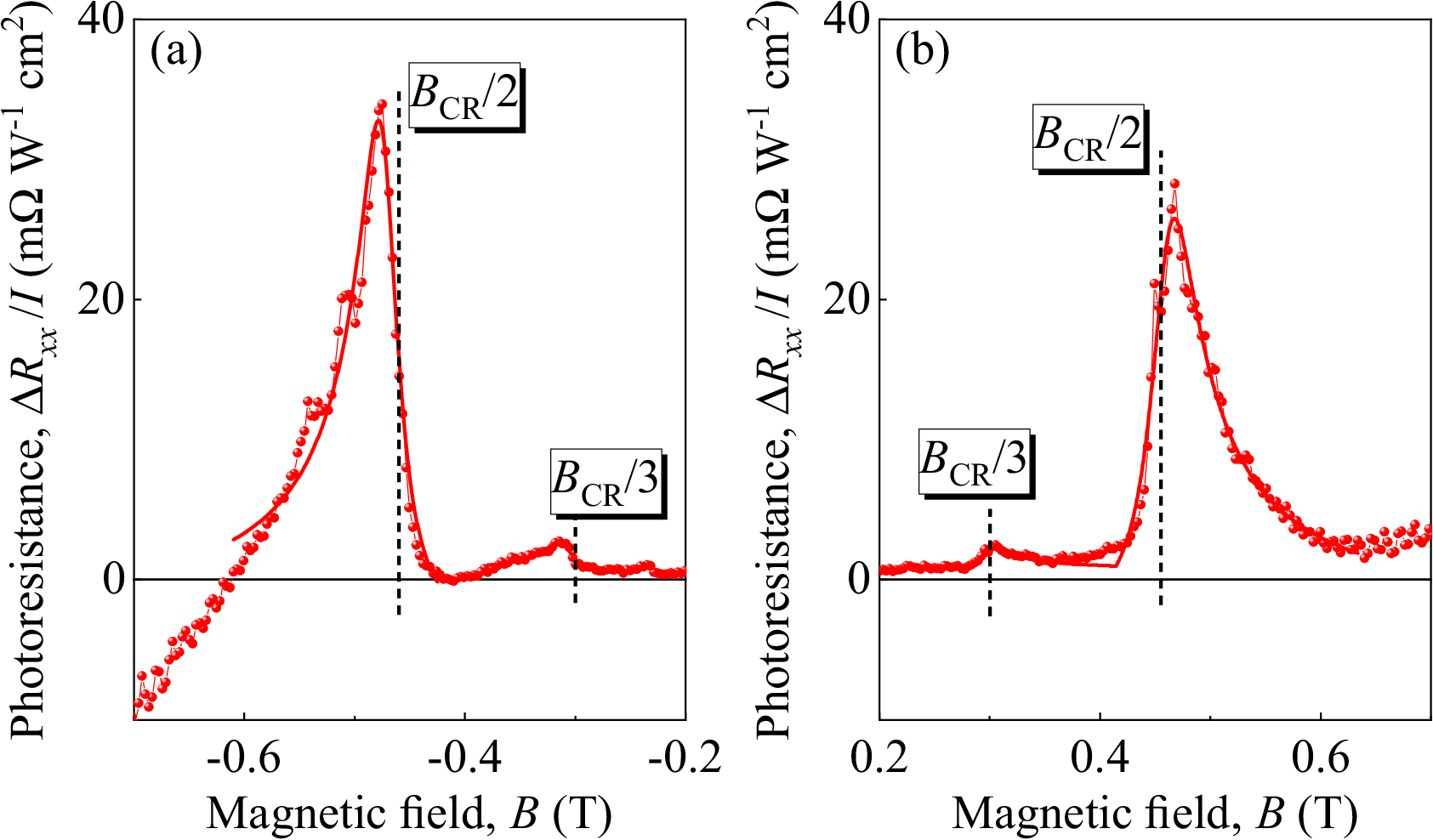}
	\caption{Graphene photoresistance (dots) measured at $T=4$ K and carrier density $n_s=3.25\times10^{12}$ cm$^{-2}$, different from that in Fig. 4 of the main text. Solid lines show the fits with model Eq. (12) of the main text. Raiation frequency $f=0.69$ THz. Panel (a) corresponds to the negative magnetic fields, while panel (b) -- to the positive ones.}
	\label{fig-extra}
\end{figure*}

The $T$-dependences of the photoresistance were recorded only for the carrier density $n_s = 3.9\times 10^{12}$ cm$^{-2}$. At lower densities, only single-temperature ($T=4$ K) $B$-dependent photoresistances were recorded. Fig.~\ref{fig-extra} shows the $\Delta R_{xx} (B)$-curve for a lower density of $3.25\times10^{12}$ cm$^{-2}$ with the corresponding fit of the data by model Eq.~(12). The width of the third CR here again appears smaller than the width of the second one. These extracted widths are shown in Fig. 4 (c) with green and orange points.

\section{Harmonic relaxation rates for impurity and phonon scattering}
We proceed to evaluate the scattering rates of angular harmonics $\gamma_m$ for the two 'conventional' scattering mechanisms, the electron-impurity and electron-acoustic phonon scattering. Assuming elastic scattering, the collision integral can be written as
\begin{equation}
	\mathcal{C}\{\delta f\} = \sum\limits_{\bf p'}{W_{\bf pp'}\delta(\varepsilon_{\bf p} - \varepsilon_{\bf p}) (\delta f_p -\delta f_{p'})},
\end{equation}
where $W_{\bf pp'}$ is the Fermi golden rule scattering probability. Plugging the angular dependence of the distribution functions in the form
\begin{equation}
	\delta f_m = \frac{\partial f_0}{\partial \varepsilon}  e^{i m \theta_p},
\end{equation}
we find the generic expression for the relaxation rate of the $m$-th angular harmonic
\begin{equation}
	\label{eq-supp-gammam}
	\gamma_m = \rho(\varepsilon_p) \int\limits_0^{2\pi}{\frac{d\alpha}{2\pi} W(p,\alpha)(1-e^{im\alpha})}.
\end{equation}
Above, we have introduced the density of states per spin and per valley $\rho(\varepsilon_p)$, and the angle between initial and final electron momenta $\alpha = \theta_p - \theta_{p'}$. At low temperatures, the momentum should be taken at the Fermi surface, $p = p_F$.

We further specify the microscopic expressions for the scattering probabilities. For electron-impurity scattering~\cite{Das_Sarma_impurities}
\begin{equation}
	\label{eq-supp-impurities-micro}
	W^{e-i}_{\bf pp'} = \frac{2\pi}{\hbar} n_{\rm imp} \left( \frac{2\pi e^2 \hbar}{|{\bf q}| +q_{TF}}\right)^2 F(q),
\end{equation}
where $n_{\rm imp}$ is the areal density of impurities, ${\bf q} = {\bf p} - {\bf p}'$ is the transferred momentum, $q_{\rm TF}$ is the Thomas-Fermi screening momentum, and the factor $F(q) = (1+\cos \alpha)/2$ accounts for the backscattering suppression and is relevant to graphene. For electron-phonon scattering in the equipartition mode (photon occupation numbers $N_B(\omega) \approx kT/\hbar\omega \gg 1$)~\cite{Vasko_Phonons}
\begin{equation}
	\label{eq-supp-phonons-micro}
	W^{e-ph}_{\bf pp'} = \frac{2\pi}{\hbar} D^2 ({\bf q}, {\bf u}_{\bf q})^2 \frac{kT}{\hbar \omega_{\bf q}} F(q),
\end{equation}
where we have introduced the deformation potential $D$ with the dimension of energy, and amplitude of zero-point atomic vibrations ${\bf u}_{\bf q} = {\bf e}_{\bf q} \sqrt{\hbar/2\rho\omega_{\bf q}}$. Introducing Eqs.~(\ref{eq-supp-impurities-micro}) and (\ref{eq-supp-phonons-micro}) into the general expression (\ref{eq-supp-gammam}) for harmonic relaxation rates, we get:
\begin{gather}
	\label{eq-supp-gammam-imp}
	\gamma^{e-i}_m = \overline{\gamma}^{e-i} \int\limits_0^{2\pi}{\frac{d\alpha}{2\pi} \frac{1-e^{im\alpha}}{(2\sin\frac{\alpha}{2} + \frac{q_{\rm TF}}{2p})^2} \times \frac{1 + \cos\alpha}{2}},\\
	\label{eq-supp-gammam-ph}
	\gamma^{e-ph}_m = \overline{\gamma}^{e-ph} \int\limits_0^{2\pi}{\frac{d\alpha}{2\pi} (1-e^{im\alpha})\times \frac{1 + \cos\alpha}{2}}.
\end{gather}
It is possible to show analytically that Coulomb relaxation rates grow linearly with $m$ in the absence of screening ($q_{\rm TF} \ll 2p$) and stay constant for strong screening ($q_{\rm TF} \gg 2p$). This result holds both for 2d systems with parabolic bands ($F(q) \equiv 1$) and for graphene, the difference between these two systems lies in the numerical prefactors. For parabolic-band 2DES 
\begin{equation}
	\gamma_{m}^{e-i}={{\bar{\gamma }}^{e-i}}\times \left\{ \begin{aligned}
		& 2m,\,\,{{q}_{\text{TF}}}\ll 2p \\ 
		& {{\left( \frac{2p}{{{q}_{TF}}} \right)}^{2}},\,\,{{q}_{\text{TF}}}\gg 2p \\ 
	\end{aligned} \right.   
\end{equation}
while for graphene
\begin{equation}
	\gamma_{m}^{e-i}={{\bar{\gamma }}^{e-i}}\times \left\{ \begin{aligned}
		& 2m-1,\,\,{{q}_{\text{TF}}}\ll 2p \\ 
		& {{\frac{1}{2} \left( \frac{2p}{{{q}_{TF}}} \right)}^{2}},\,\,{{q}_{\text{TF}}}\gg 2p \\ 
	\end{aligned} \right.   
\end{equation}
Both dependences are illustrated in Fig.~\ref{fig_imp_harmonics}.
\begin{figure*}[h]
	\center{\includegraphics[width=0.9\linewidth]{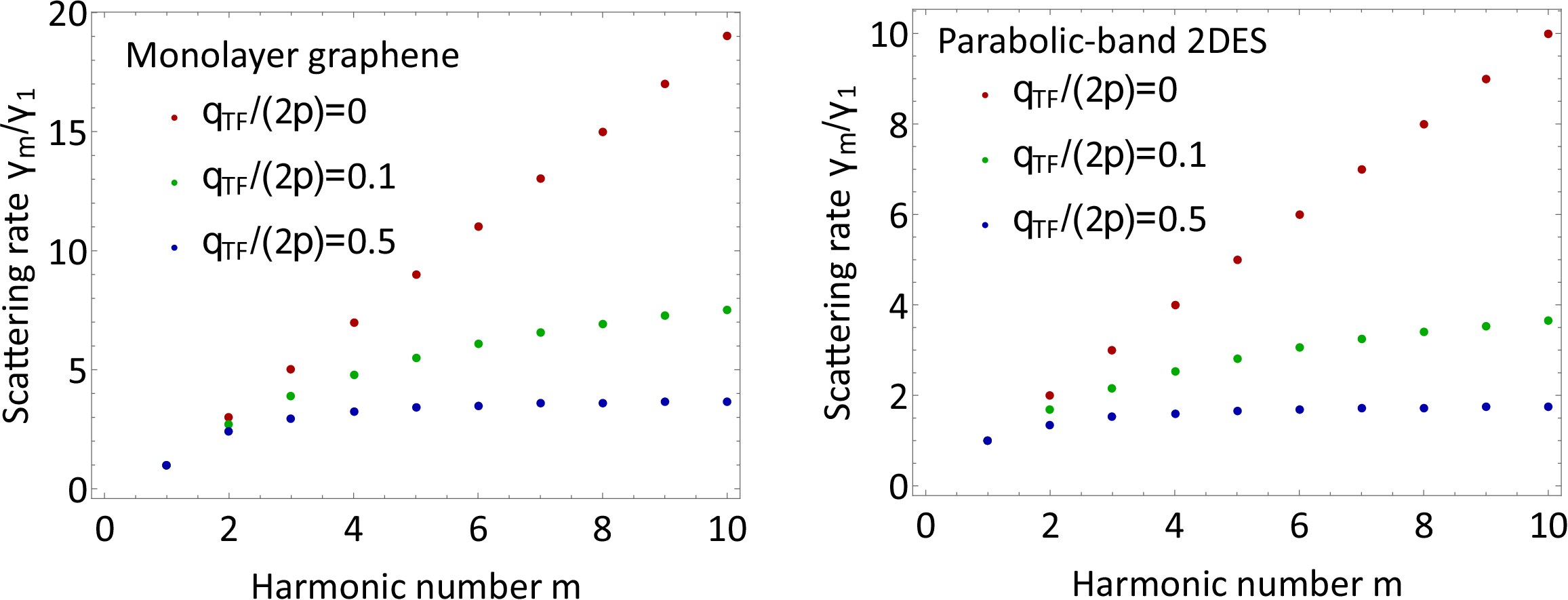}}
	\caption{Relaxation rates of various angular harmonics $\gamma_m$ (normalized by momentum relaxation rate $\gamma_1$) for Coulomb scattering by charged impurities. Left panel corresponds to chiral electrons in monolayer graphene, right panel -- to the massive electrons in parabolic-band 2DES. Different colors correspond to different screening strengths parameterized by the ratio $q_{\rm TF}/2p$ 
	}
	\label{fig_imp_harmonics}
\end{figure*}

For electron-phonon scattering, the harmonic relaxation rates do not depend on $m$ at all. This applies both to parabolic-band 2DES, where
\begin{equation}
	\gamma_{m}^{e-ph}={{\bar{\gamma }}^{e-ph}},
\end{equation}
and to graphene, where
\begin{equation}
	\gamma_{m}^{e-ph}=\frac{1}{2} {{\bar{\gamma }}^{e-ph}}.
\end{equation}

The phonon equipartition assumed upon derivation of $\gamma_{m}^{e-ph}$ is, strictly speaking, not realized in our low-temperature experiment. It is possible to show that in the opposite (Bloch - Gruneisen (BG) limit) $\gamma_m$ grow rapidly with increasing $m$. Indeed, electron scattering in the BG limit occurs in small-angle steps $\alpha \sim s/v_F$. Such small-angle scattering cannot relax the smooth distribution function harmonics ($m \sim 1$), but efficiently relaxes the rapidly-varying distribution function harmonics ($m \gg 1$).

To conclude, we observe that electron-impurity and electron-phonon scattering produce relaxation rates that never decay  with increasing $m$. This contrasts to electron-electron scattering in the tomographic Fermi liquid, where even-mode relaxation rates are faster than odd-mode relaxation rates.

\end{widetext}

\bibliography{References}

\maketitle

\end{document}